# A Deep Learning Approach for Predicting Two-dimensional Soil Consolidation Using Physics-Informed Neural Networks (PINN)

Yue Lu, Gang Mei*, Francesco Piccialli

School of Engineering and Technology, China University of Geosciences (Beijing)

Department of Mathematics and Applications, University of Naples Federico II, Italy

Corresponding author: gang.mei@cugb.edu.cn

## Abstract

Soil consolidation is closely related to seepage, stability, and settlement of geotechnical buildings and foundations, and directly affects the use and safety of superstructures. Nowadays, the unidirectional consolidation theory of soils is widely used in certain conditions and approximate calculations. The multi-directional theory of soil consolidation is more reasonable than the unidirectional theory in practical applications, but it is much more complicated in terms of index determination and solution. To address the above problem, in this paper, we propose a deep learning method using physics-informed neural networks (PINN) to predict the excess pore water pressure of two-dimensional soil consolidation. In the proposed method, (1) a fully connected neural network is constructed, (2) the computational domain, partial differential equation (PDE), and constraints are defined to generate data for model training, and (3) the PDE of two-dimensional soil consolidation and the model of the neural network is connected to reduce the loss of the model. The effectiveness of the proposed method is verified by comparison with the numerical solution of PDE for two-dimensional consolidation. Using this method, the excess pore water pressure could be predicted simply and efficiently. In addition, the method was applied to predict the soil excess pore water pressure in the foundation in a real case at Tianjin port, China. The proposed deep learning approach can be used to investigate the large and complex multi-directional soil consolidation.

# 1. Introduction

Soil deformation and stability problems associated with soil consolidation are faced in the construction of large infrastructures such as highways, embankments, and airports. Soil consolidation laws are complex and depend not only on the type and properties of the soil but also on its boundary conditions, drainage conditions, and types of loading. Therefore, to ensure the safety of infrastructures, the study of multi-directional soil consolidation theory, which is closer to the actual working conditions, has broad application prospects and economic value.

There is much research work on soil consolidation. Terzaghi [1] presented his classical consolidation theory of soil in 1925 as a symbol of the birth of soil mechanics, which became an important milestone in the history of soil mechanics. Biot [2] proposed the Biot consolidation theory based on the effective stress principle, soil continuity, and equilibrium equation under the condition of considering the relationship between pore pressure and skeleton deformation during soil consolidation. Schiffman [3] investigated the consolidation equation for the case of linear increase of load with time and presented an analytical solution for one-dimensional soil consolidation under this situation. Wilson & Elgoharhy [4] analyzed the one-dimensional consolidation of clay soils under cyclic loading. Zhu et al. [5] investigated the one-dimensional consolidation of double-layered foundations under variable loads varying in-depth. Indraratna, B. [6] proposed a method for radial consolidation of clays using compression index and varying horizontal permeability. Lo [7] studied the consolidation and derivation of pore water and air pressure and total sedimentation of one-dimensional unsaturated soil. Pham et al. [8] proposed a new method to predict the consolidation coefficients of soft ground foundations and verified the accuracy of the prediction by soil samples collected from the project site. Yarushina [9] established the theory of nonlinear viscoelasticity of porous materials under gravity and expressed the compressive coefficients of low porosity viscoelastic materials with constitutive equations.

In regards to the drainage boundary condition problem in the one-dimensional consolidation, Ho [10] used the feature equation expands and Laplace transforms techniques, the precise solution of the one-dimensional consolidation flow equation for unsaturated soil layer was found, and the solution under two different boundary conditions is elaborated. Cavalcanti [11] discussed the application of boundary elements with time-independent fundamental solutions in solving the plane strain consolidation equation for porosity-elastic saturated media.

Moreover, Joaquim [12] performed a numerical simulation of fluid flow in porous media according to Darcy's law, considering the stress distribution between fluid and pore space according to the Terzaghi effective stress principle and approximating the partial differential equation (PDE) obtained by using the finite difference method of mesh discretization solution. Wang et al. [13] derived a semi-analytic solution for unsaturated soils with semi-permeable boundary conditions.

In summary, it is easy to learn that most research work focuses on soil consolidation on one-dimensional drainage consolidation, while little research has been conducted on two-dimensional or three-dimensional soil consolidation. The consolidation problems in practical engineering are generally complex, and general foundations always cause drainage and deformation in multiple directions when subjected to building loads. This leads to the result that the settlement rate calculated by unidirectional consolidation theory is slower than what occurs.

Currently, there are two problems when analyzing two-dimensional or three-dimensional soil consolidation. (1) The modeling process of traditional numerical methods is quite complicated for high-dimensional problems. (2) In general, the traditional numerical methods are computationally quite inefficient when investigating multi-dimensional soil consolidation.

To address the above problems, in this paper, we propose a deep learning method to predict two-dimensional soil consolidation using PINNs. In the proposed method, the prediction of excess pore water pressure is demonstrated for different boundary conditions: drainage at the top boundary and drainage at the top and bottom boundary. First, we use DeepXDE[14], a library in Python, to define the computational domain, PDEs, constraints, and training data for two-dimensional soil consolidation, then construct the neural network, and finally, we connect the PDE of two-dimensional soil consolidation and the model of the neural network to reduce the loss of the model. Using this method, the excess pore water pressure of soil could be predicted simply and efficiently.

Physics-informed neural network (PINN) is a type of neural network for solving PDEs using physical equations as operational constraints[15]. The most essential idea behind the PINN is to convert physical constraints as the additional loss functions in deep neural networks[16]. More details about the PINN will be introduced in Section 2.2.

The rest of this paper is organized as follows. Section 2 describes the details of this proposed method. Section 3 verifies this proposed method in two simple examples and analyses the results. In Section 4, the proposed method is applied to predict the consolidation of foundation soils in a real case at Tianjin port, China. Section 5 discusses the advantages and shortcomings of the proposed deep learning method and points out future work. Finally, Section 6 concludes the paper.

## 2. Methods
### 2.1 Overview of the Proposed deep learning method

In this paper, we propose a deep learning approach using PINN to predict the excess pore water pressure of two-dimensional soil consolidation (see Fig.1). First, we construct a fully connected neural network. Second, we define the PDE, time domain, and initial and boundary conditions for two-dimensional soil consolidation in the DeepXDE, a Python library. Third, we connect the PDE to the neural network and tune the parameters to reduce the model loss. Finally, we employ the trained model to predict the excess pore water pressure. We verify this proposed method in two simple examples: two-dimensional consolidation for drainage at the top boundary and drainage at the top and bottom boundary.

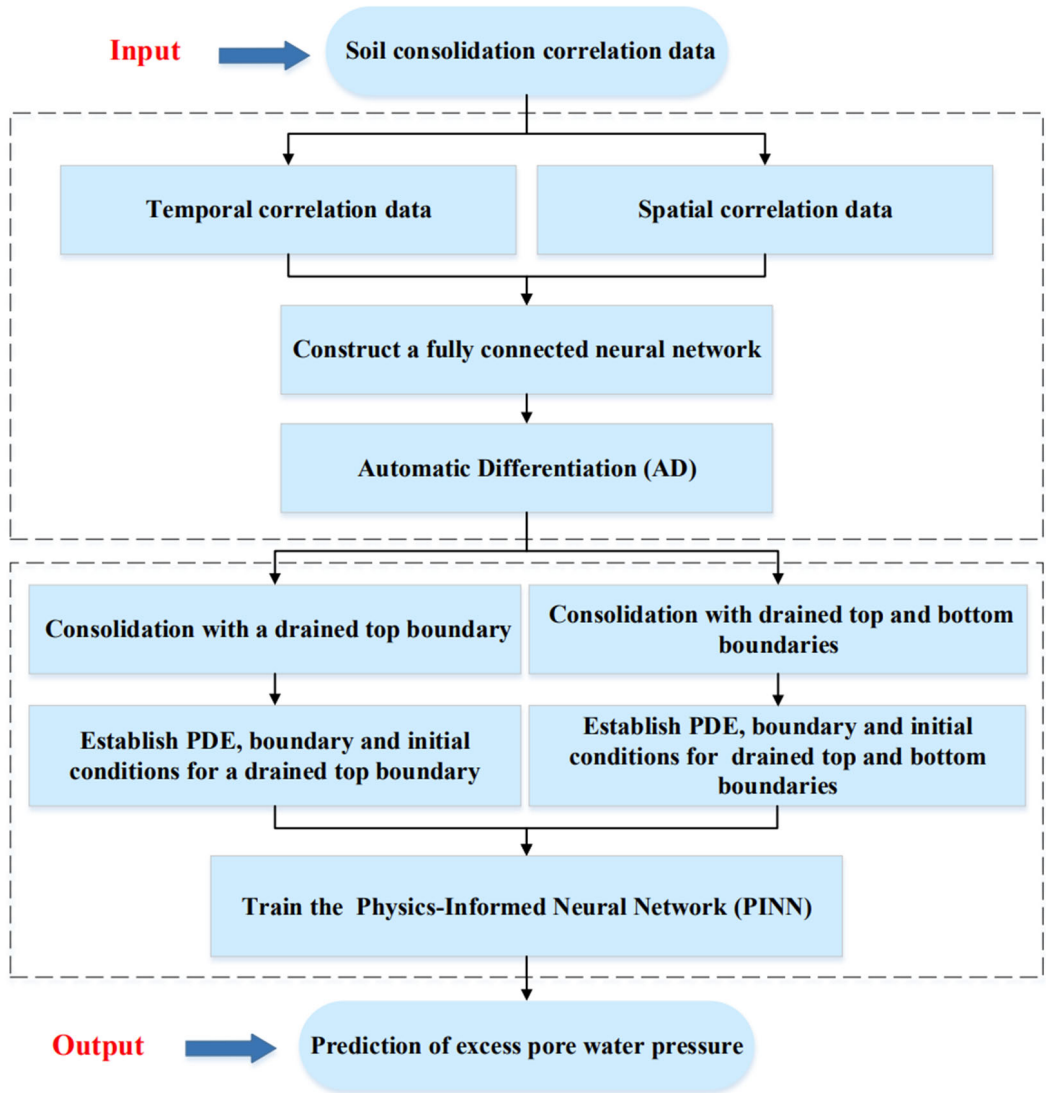

**Fig.1** Flowchart of the proposed deep learning method.

**2.2 Background and Theory of PINN**

In this section, we will introduce how to employ PINN to solve the PDE.

**2.2.1 Background of PINN**

The idea of applying prior knowledge to deep learning was first proposed by Owhadi[17] in 2015. Subsequently, Raissi et al.[18,19] used Gaussian process regression to establish a representation of linear operator generalization to present uncertainty estimates for various physical problems, introducing and illustrating the PINN method for solving nonlinear PDEs[15]. Karniadakis et al.[20] proposed physics-informed machine learning as an algorithm that combines incomplete data with physical prior knowledge and discussed its various applications in forward and inverse problems.

Currently, PINN has been increasingly used in various engineering problems such as fluid mechanics [21-24]. For example, Bandai, T. et al.[25] proposed the constitutive relation and soil water flux density for volumetric water content measurement based on physical information neural network. Zhang, Z.[26] used the physics-informed neural network to simulate and predict the transient Darcy flow of unlabeled data in heterogeneous reservoirs. In addition, Bekele, Y. W.[27] used PINN to solve forward and inverse problems of one-dimensional consolidation of soils.

**2.2.2 Theory of PINN**

PINN combines PDEs and physics-informed constraints into the computation of loss function to constrain the neural network and reduce the training loss, replacing the actual observed data of the model, i.e., the "data-free" neural network. It approximates the PDE solution by training the neural network to minimize the loss function, including terms along the boundary of the space-time domain reflecting the initial and boundary conditions, and residuals of PDEs at selected points in the domain. The values in the input domain, by combining this with physical information, generate an estimated solution to the point differential equation after training.

The process of solving the PDE requires the derivative of the input values. There are four methods for calculating derivatives: hand-coded, symbolic, numerical, and automatic. However, it is impractical to calculate the derivatives manually in the face of complex equations. The automatic differentiation (AD) used in PINN, use exact expressions with floating-point values rather than symbolic strings, there is no approximation error[28]. There is no doubt that the prediction accuracy and efficiency have been improved.

PINN is composed of physical information, neural networks, and feedback mechanisms[29]. First, the physics-informed is used to calculate the partial derivatives of the functions and to determine the loss of the equation terms. Then, the model is started to be trained by connecting the two modular neural networks through a differentiation algorithm. Finally, continuous feedback adjustments are made to minimize the training losses. The PINN workflow schematic is illustrated in Fig.2.

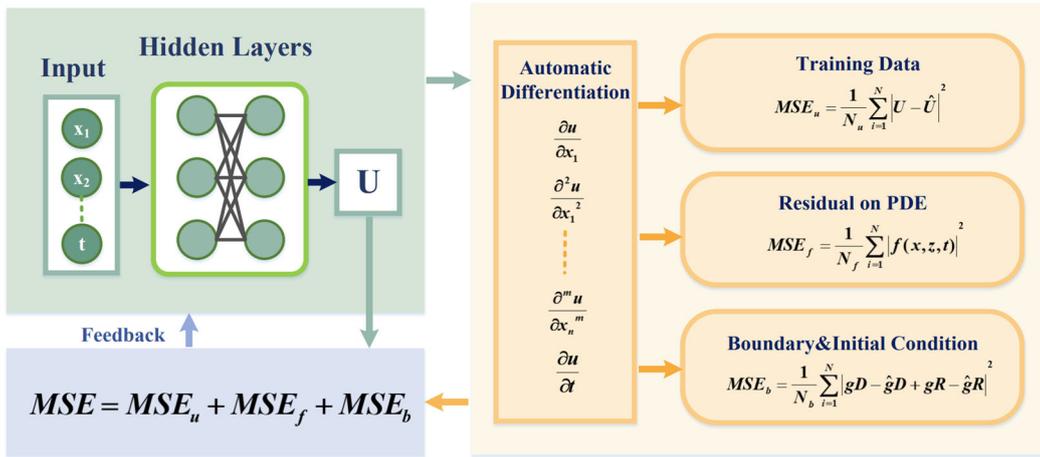

**Fig.2** Illustration of a PINN algorithm for solving partial differential equations. This method uses AD technology to analyze and derive the integer derivative, and the obtained MSE is fed back to the neural network

Physics-informed neural network learning by minimizing the loss of mean squared error. The mean square error of this neural network model is displayed in Eq. (1).

$$MSE = MSE_u + MSE_f + MSE_b \qquad (1)$$

where

$$MSE_u = \frac{1}{N}\sum_{i=1}^{N}|u(x,z,t) - \hat{u}(x,z,t)|^2$$

$$MSE_f = \frac{1}{N_f}\sum_{i=1}^{N_f}|f(x,z,t)|^2$$

$$MSE_b = \frac{1}{N_b}\sum_{i=1}^{N_b}|g_D(x,z,t) - \hat{g}_D(x,z,t) + g_R(x,z,t) - \hat{g}_R(x,z,t)|^2$$

Here, (x, z, t) is the input to the training of a neural network model. In the proposed method, training points are randomly generated based on the physical constraints of the governing PDE. $g_D$ (x, z, t) represents the initial training points, $g_R$ (x, z, t) represents the boundary training points.

In this paper, we use the Python library DeepXDE[14] to solve practical applications with PINNs. Solving differential equations with DeepXDE uses built-in modules to specify problems, including computational domains (geometry and time), PDEs, boundary/initial conditions, and neural network architecture[14]. The workflow of DeepXDE is shown in Fig.3. Furthermore, four boundary conditions, including Dirichlet, Neumann, Robin, and periodic are provided by this library. Initial conditions can be defined by IC modules. Such as loss type, metric, optimizer, learning rate table, initialization and regularization can be adjusted and selected by yourself according to different needs.

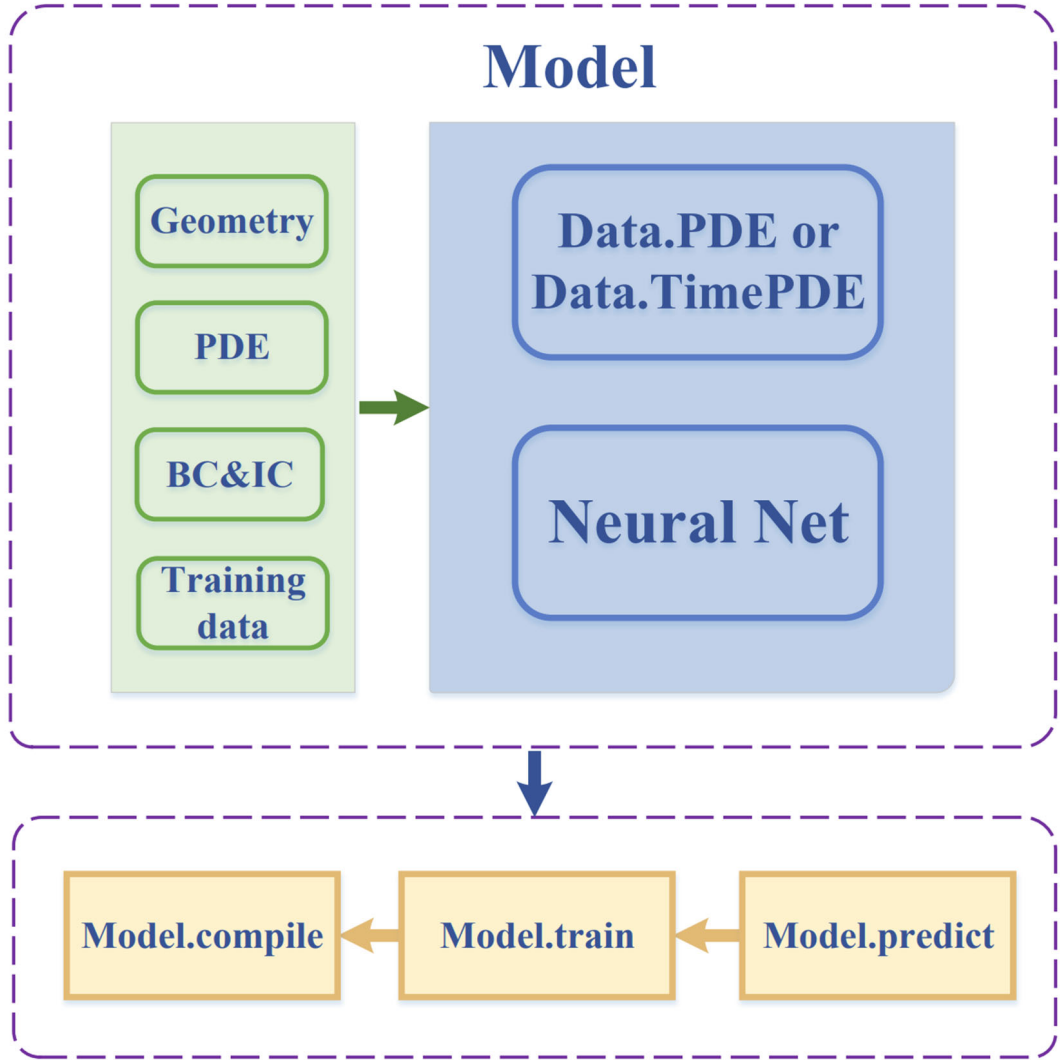

**Fig.3** The workflow of DeepXDE. The green modules define the PDE and the training hyperparameters. The blue modules combine the PDE and training hyperparameters. The yellow modules are the three steps to solve the PDE.

**2.3 Problem 1: Two-dimensional Soil Consolidation for Drained at the Top Boundary**

In this section, we will introduce how to employ the deep learning approach to solve the problem of two-dimensional consolidation for drained at a top boundary.

Rendulic[30] extended the one-dimensional consolidation theory to two or three dimensions and proposed the Terzaghi-Rendulic theory, assuming that the sum of normal stresses at any point in soil under constant external loads is a constant in consolidation. Therefore, the consolidation problem is the same as the thermal diffusion problem of consolidation, and its mathematical expression is also called the diffusion equation (see Eq. (2)).

$$\frac{\partial u}{\partial t} - C_V \left( \frac{\partial^2 u}{\partial^2 x} + \frac{\partial^2 u}{\partial^2 z} \right) = 0 \qquad (2)$$

where $u$ represents the excess pore water pressure, $Cv$ represents the soil consolidation coefficient, and x and z represent the horizontal and vertical directions of soil layer, respectively.

For the two-dimensional consolidation of drainage at the top boundary, it is assumed that the bottom boundary is impervious. The excess pore water pressure dissipates only at the top boundary. The top boundary satisfies the Dirichlet boundary condition: u(x) = 0, and the bottom boundary satisfies the Neumann boundary conditions: $\frac{\partial u}{\partial z} = 0$. A schematic diagram of consolidation for drained at a top boundary is displayed in Fig.4. Assume that the initial excess pore water pressure distribution is q, and the initial excess pore pressure is uniformly distributed and equal to the surface overload. In addition, we set the thickness of the soil layer as H. The boundary conditions are mathematically expressed as Eq. (3).

$$\begin{cases} u = 0 & at \Gamma_b \quad t > 0 \\ \frac{\partial u}{\partial z} = 0 & at \Gamma_t \quad t > 0 \end{cases} \tag{3}$$

In the proposed method, we use the PDE, boundary, and initial conditions of consolidation for drained at a top boundary to generate training data, and then the trained model is applied to predict the excess pore water pressure.

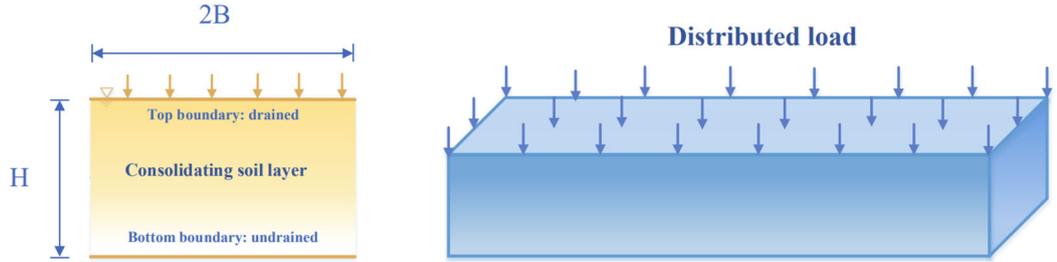

**Fig. 4** Schematic diagram of two-dimensional soil consolidation for drainage at the top boundary.

## 2.4 Problem 2: Two-dimensional Soil Consolidation for Drained at the Top and Bottom Boundaries

In this section, we will introduce how to employ the deep learning approach to solve the problem of two-dimensional soil consolidation for drained at the top and bottom boundaries.

When both the top and bottom of the foundation are drainable boundaries, the excess pore water pressure is dissipated by both boundaries. Top and bottom boundaries satisfy the Dirichlet boundary condition: u(x) = 0. A schematic diagram of consolidation drained at the top and bottom boundaries is displayed in Fig.5. In mathematics, this condition of soil consolidation for drained at top and bottom boundaries is expressed as Eq. (4).

$$u = 0 \quad at \Gamma_t \cup \Gamma_b, t > 0 \tag{4}$$

Similarly, we set the initial excess pore pressure distribution as q. However, the maximum drainage distance for consolidation for drained at top and bottom boundaries is taken as half the thickness of the soil layer. Therefore, the thickness of the soil layer is set to double of drained at the top boundary i.e.,2H. We use the PDE, boundary, and initial conditions of it to generate training data, then the trained model is applied to predict the excess pore water pressure.

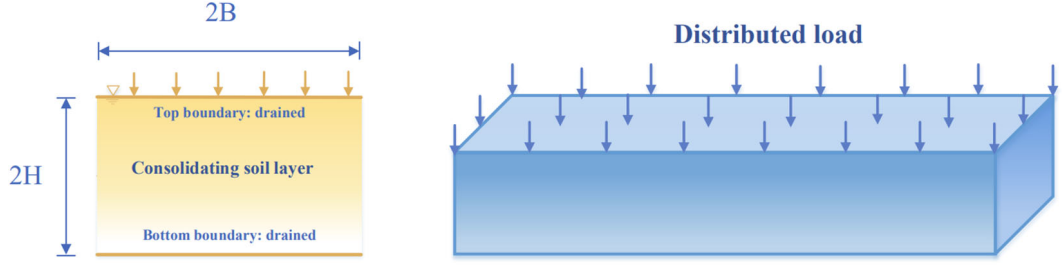

**Fig. 5** Schematic diagram of two-dimensional soil consolidation for drained at the top and bottom boundaries.

### 3. Validation of the Proposed Deep Learning Approach

In this section, to verify the effectiveness of the proposed deep learning approach, we apply this proposed method with simple data to different boundary conditions and compare the results predicted by our approach with the numerical solutions obtained by the improved weighted residual method.

### 3.1 Experimental Environment

The experiments are conducted on a laptop with an NVIDIA GeForce RTX3070 Laptop GPU and an AMD Ryzen 7 5800H with Radeon Graphics. Furthermore, the library used to implement the PINN is DeepXDE version 0.13.6.

### 3.2 Results of Consolidation with A Drained Top Boundary

According to the weighted residual method, we assume that there is drainage sand well at the center of the substrate (load) and consider its influence range in the horizontal direction to be limited, which is shown by the numerical analysis as A=2B. Since there is usually no drainage sand well at the center of the load in actual projects, the numerical solution is obtained after excluding this condition (Eq. (5)).

$$u = \frac{16q}{\pi^2} \sum_{m=1,3,5\ldots}^{\infty} \frac{1}{m^2} \sin(\frac{m\pi}{2A}(x+A))\sin(\frac{m\pi}{2H}z)e^{-\left(\frac{1}{A}+\frac{1}{4H^2}\right)^2 c_v m^2 \pi^2 t} \quad (5)$$

For a numerical example of a drained top boundary, we set the soil layer thickness to H=1m, the soil consolidation coefficient is $C_v = 0.01$ cm$^2$/s. In addition, we assume that the foundation is subjected to distributed load $q = 5$ KN/m$^2$ and that the load level affects the range A = 1m. The numerical solution Eq. (5) is used as the reference solution for the training results. The geometry module of this example is Rectangle [-1,0] [1,1]. Soil consolidation is a time-dependent PDE problem, And the time domain calculated in this experiment is from 0 to 1. Finally, the input of the PDE system and the construction of the physical information model are completed. The residuals were tested by sampling 100 points in the domain, initial and boundary conditions, and using 1000 points to test the PDE residuals.

Here, we used a fully connected neural network of depth 6 (i.e.,5 hidden layers) and width 32. For temporal and spatial partial derivatives of excess pore water pressure, this was achieved by AD in this neural network. Values of (x, z, t) are used as the input of the neural network, where this model predicts the excess pore water pressure as the output. Adam optimizer was chosen to train 10,000 epochs for this experiment, and the time required was approximately 15 seconds. The time spent on model training is proportional to the number of hidden units, hidden layers, and training epochs, we tune the parameters depending on the desired accuracy.

On the established soil consolidation for drained at a top boundary model, 10 points are randomly selected, the numerical solution and the predictive solution are entered, and the color maps of the numerical solution and the predictive solution at different times are obtained by interpolation shown in Fig.6.

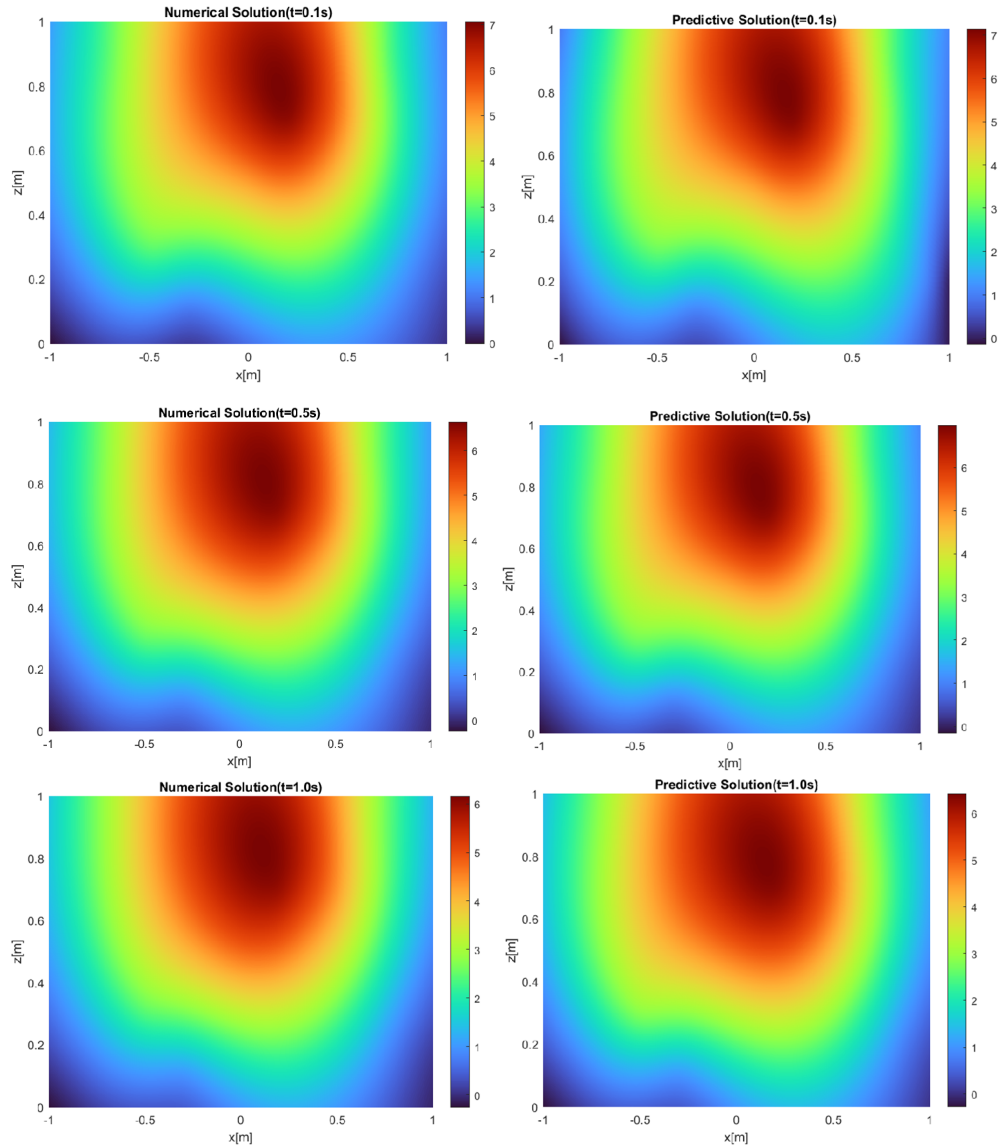

**Fig.6**. Color maps for numerical solution and predicted solution of two-dimensional soil consolidation with a drained top boundary at different times.

The final train loss and test loss of this two-dimensional soil consolidation for drained at a top boundary model are displayed in Fig. 7 and Fig.8.

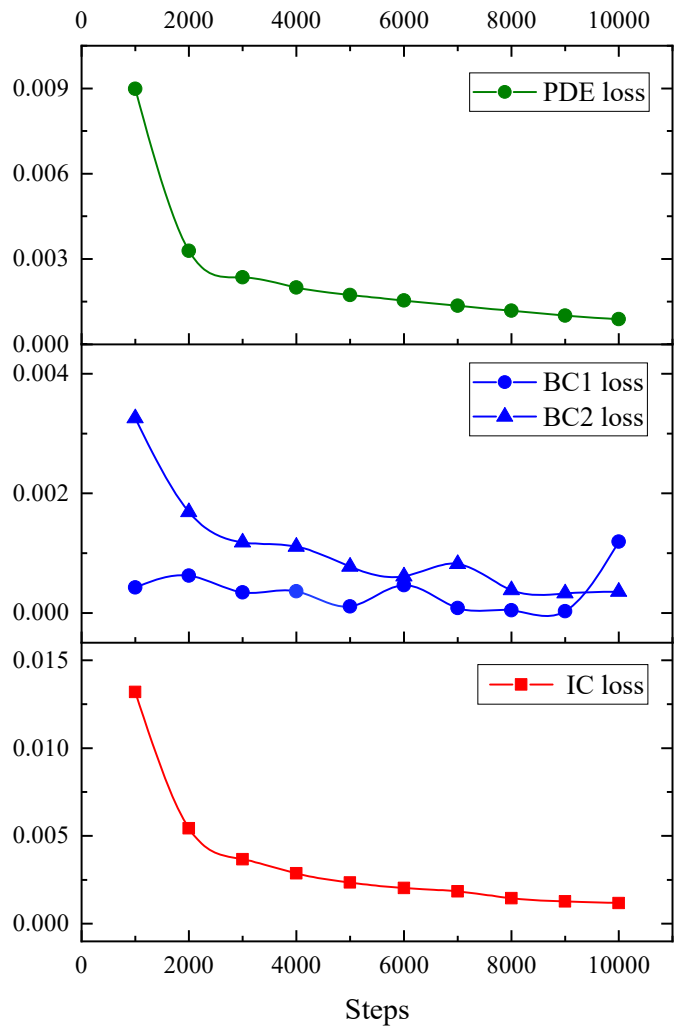

**Fig.7** Training results for two-dimensional consolidation model of drained at the top boundary.

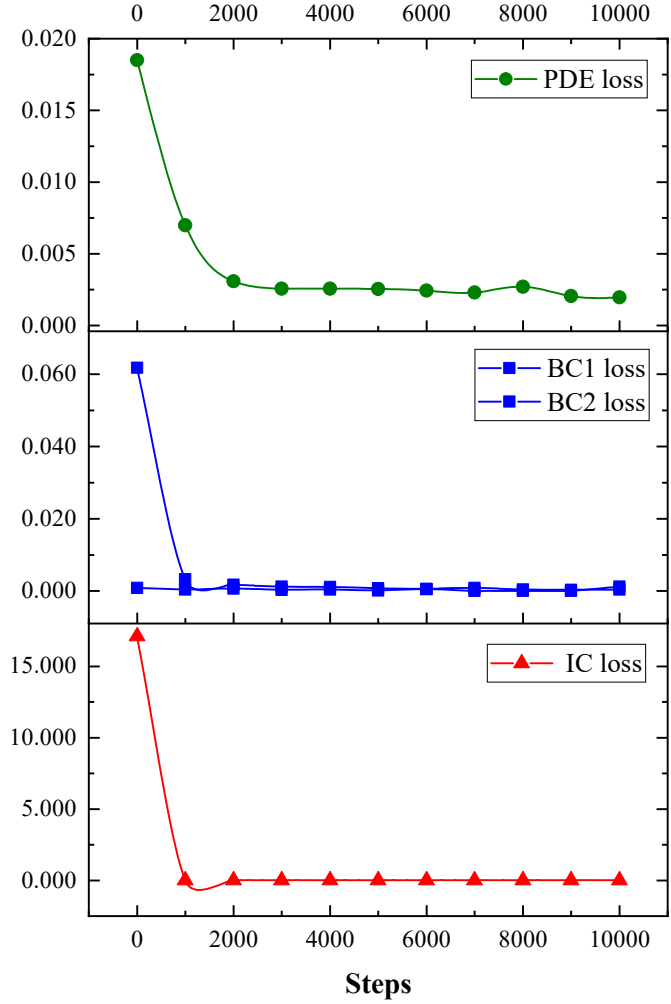

**Fig.8** Testing results for two-dimensional consolidation model of drained at the top boundary.

According to Fig.7 and Fig. 8, it is observed that the loss of PDE, boundary, and initial conditions of this model in training and testing has a good downward trend and gradually tends to be stable after 2000 epochs of training.

The test measure in this experiment is the ratio of training loss to the numerical solution, which better reflects the training results of the model (see Eq. (6)). The final mean squared error loss and test metric of a drained top boundary model with two-dimensional consolidation are displayed in Fig.9.

$$test\ metric = \frac{u-\hat{u}}{u} \qquad (6)$$

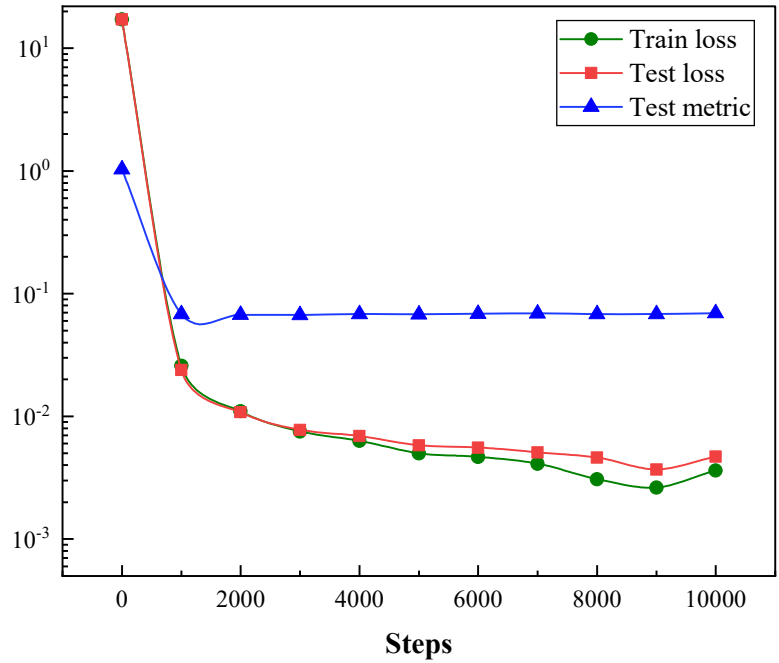

**Fig.9** Mean squared error loss and test metric for two-dimensional consolidation model of drained at the top boundary.

In this case, The training loss drops to $2.64 \times 10^{-3}$, the test loss drops to $3.67 \times 10^{-3}$, and the test metric drops to $6.82 \times 10^{-2}$.

**3.3 Results of Consolidation for Drained at the Top and Bottom Boundaries**

For two-dimensional consolidation drained at the top and bottom boundaries, we contemplate using the same neural network model as described in Section 3.2. However, the excess pore water pressure in this condition is permitted to dissipate through both boundaries, corresponding to the absence of pore water in the center of the soil layer. For comparison, we set the soil layer thickness to be 2H = 2m, the geometry module of this example is Rectangle [-1,0] [1,2]. The other constraints are the same as Section 3.2.

Similarly, we used a fully connected neural network of depth 6 (i.e., 5 hidden layers) and width 32. For temporal and spatial partial derivatives of excess pore water pressure, this is achieved by AD in this neural network. Values of (x, z, t) are used as the input of the neural network, where this model predicts the excess pore water pressure as the output. Adam optimizer was chosen to train 10,000 epochs for this experiment. Since the boundary constraints of double-sided drainage are simpler than those of single-sided, the training time of the model is shorter.

On the established soil consolidation for drained at top and bottom boundaries model, 10 points are randomly selected, the numerical solution and the predictive solution are entered, and the color maps of the numerical solution and the predictive solution at different times are obtained by interpolation, as displayed in Fig. 10 below.

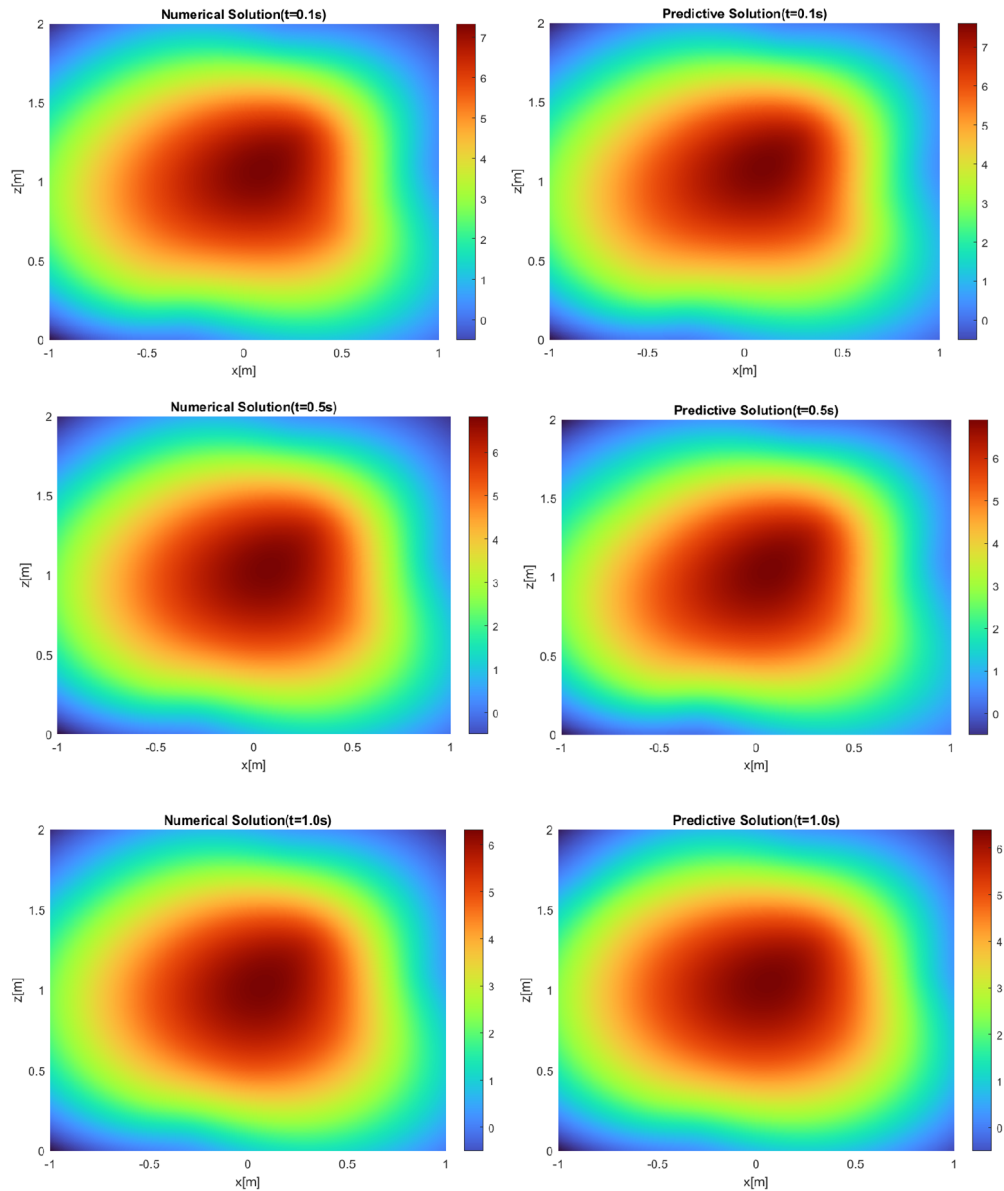

**Fig.10** Color maps for numerical solution and predicted solution of two-dimensional soil consolidation for drained at the top and bottom boundaries at different times.

The final train loss and test loss of this two-dimensional soil consolidation for drained at the top and bottom boundaries model are displayed in Fig .11 and Fig .12.

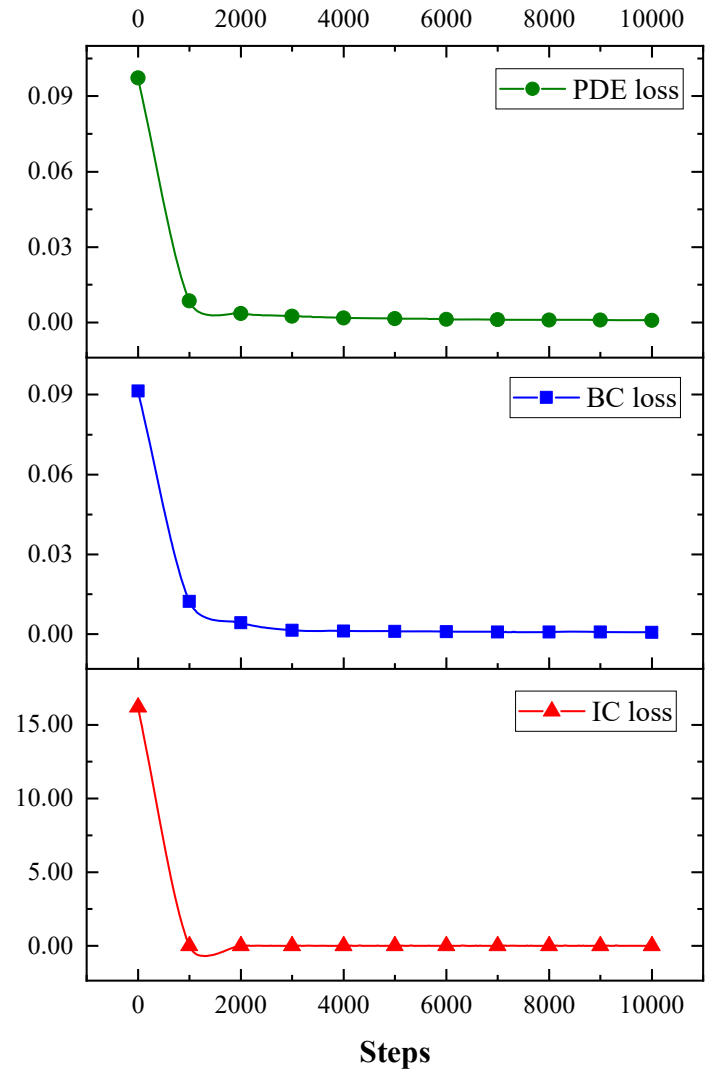

**Fig.11** Training results for two-dimensional consolidation model of drained at the top and bottom boundaries.

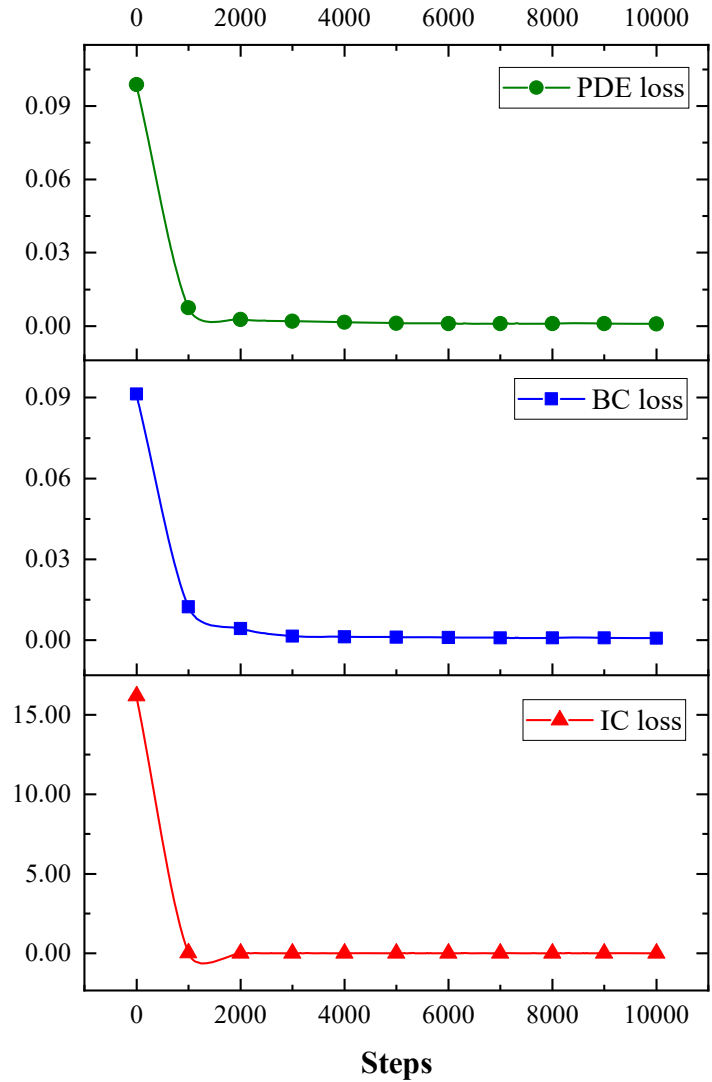

**Fig.12** Testing results for two-dimensional consolidation model of drained at the top and bottom boundaries.

According to Fig.11 and Fig.12, it is observed that the loss of PDE, boundary and initial conditions of this model in training and testing have a good downward trend and gradually tend to be stable after 2000 epochs of training.

The final mean squared error loss and test metric of two-dimensional consolidation for drained at the top and bottom boundaries are displayed in Fig. 13.

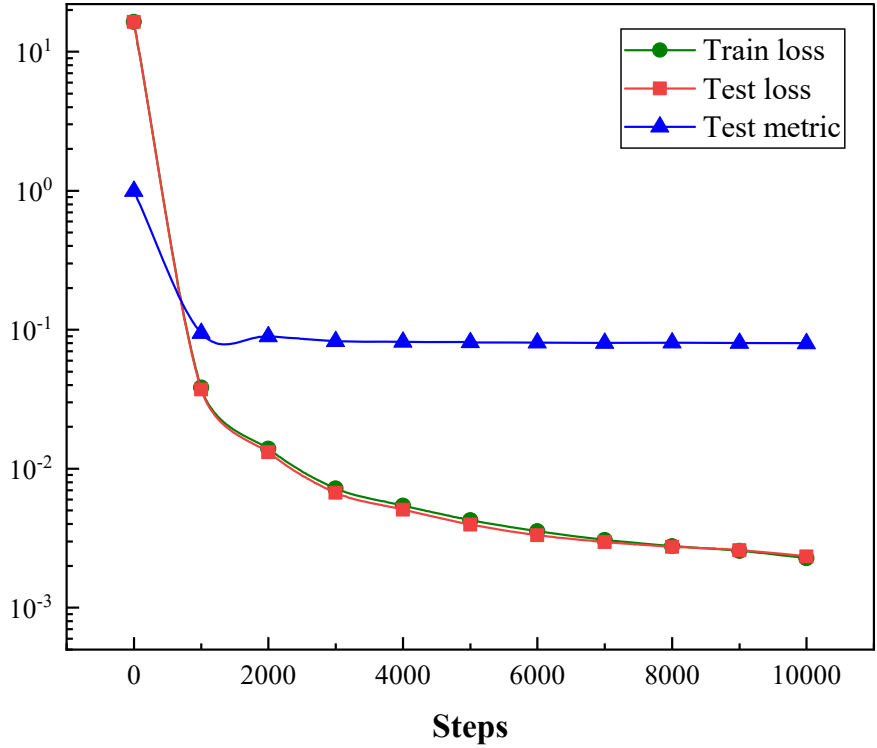

**Fig.13** Mean squared error loss and test metric for two-dimensional consolidation model of drained at the top and bottom boundaries.

In this case, the training loss drops to $2.27\times10^{-3}$, the test loss drops to $2.34\times10^{-3}$, and the test metric drops to $7.96\times10^{-2}$. We observed the great performance of PINN models in predicting excess pore water pressure.

## 4. Application of the Proposed Deep Learning Approach

In this section, the proposed prediction approach is used to predict the time required to complete the settlement consolidation of foundation soil in this case study area.

### 4.1. Engineering Background

The study area of the application case is derived from reference [31]. In our application case, we directly use the engineering background information to build the computational model and then use the proposed PINN-based method to predict the excess pore water pressure in this foundation.

A section of the road from the port of Tianjin, China, needed to be built in the soil thickness is 20 m of soft clay layer [31]. Before proceeding with the construction of the road, the soil in the area needed to be improved and loads applied to the foundation to minimize excess pore water pressure.

The road section is 364.5 m long and 51 m wide. This foundation with a soil thickness of 20 m is roughly divided into three layers, the top layer is clay slurry and silt with a thickness of 8 m, and the consolidation coefficient is $2.25\times10^{-3}$ cm$^2$/s. The middle layer is soft clay with a thickness of 8 m, the consolidation coefficient is $2.59\times10^{-3}$ cm$^2$/s. And the bottom layer is silty clay with a thickness of 4 m, the consolidation coefficient is $2.68\times10^{-3}$ cm$^2$/s. The schematic diagram is shown in Fig.14.

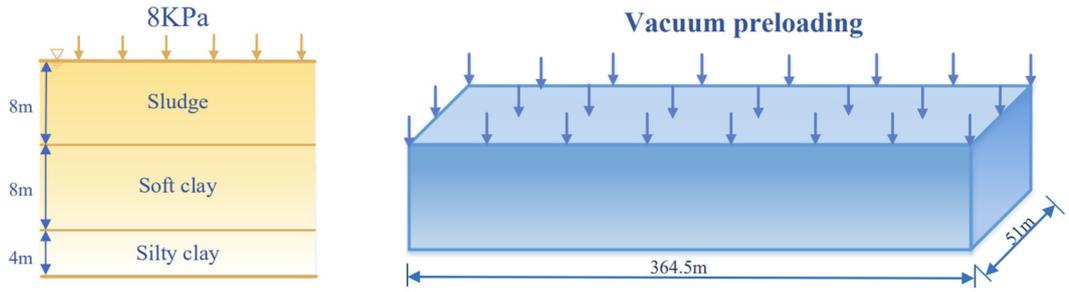

**Fig.14** Schematic diagram of the foundation soil layer of the road section.

### 4.2 Predictive Model

In this section, the proposed PINN approach is employed to predict the excess pore water pressure of the foundation in this case study area. The bottom of the foundation is the impervious bed, therefore, road improvements to the port are a problem of consolidation for drained at the top boundary, and this model is constructed in the same way as Section 2.3.

To advance the shear strength of the foundation soil layer, a vacuum load of 80 KPa is applied to the foundation. And the load level affects the range A = 182.5m, the thickness of the foundation soil layer H=20m. Therefore, the geometry module of this example is Rectangle [-182.5,0] [182.5,20]. Three different prediction models were developed based on the consolidation coefficients of different soil layers. The residuals were tested by sampling 1000 points in the domain, initial and boundary conditions, and using 1000 points to test the PDE residuals. Similarly, a fully connected neural network of depth 6 (i.e.,5 hidden layers) and width 32 is used. The Adam optimizer was chosen to train 50,000 epochs for this experiment.

### 4.3 Predictive Results

To obtain the time for soil consolidation to complete, i.e., the time required for the excess pore water pressure to completely dissipate. Different time points are selected in turn, and after the training of the constructed model in Section 4.2, color maps of the distribution of excess pore water pressure at different time points in this road section foundation can be obtained as shown in Fig.15.

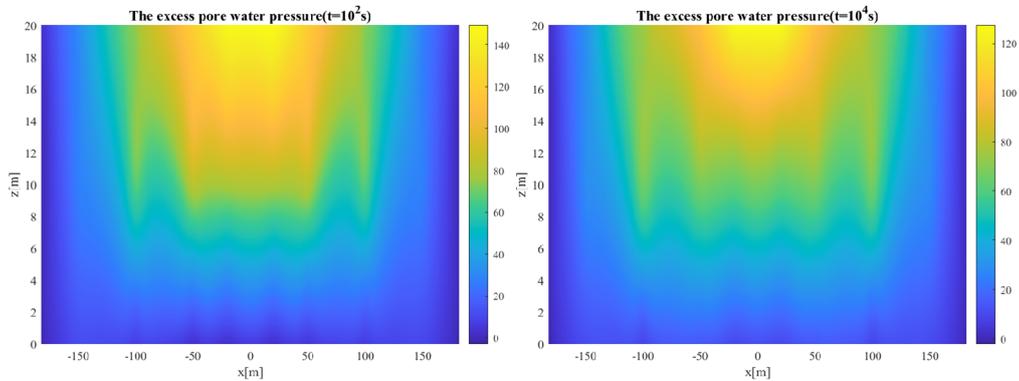

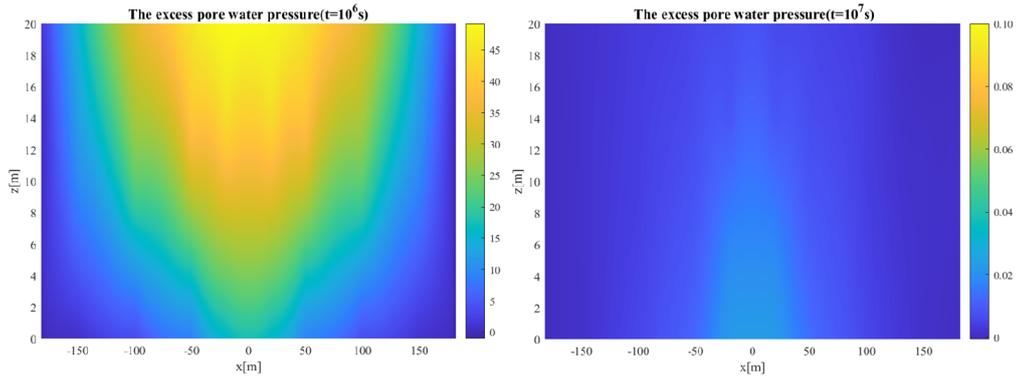

**Fig.15** Color plot of excess pore water pressure dissipation with time.

According to Figure 15, it is observed that the excess pore water pressure in the foundation soil layer progressively dissipates at any time after the load is applied to the foundation pavement with time. Due to the impermeable layer at the bottom of the foundation of this port section, the excess pore water pressure seeps out from both sides of the soil layer as well as from the top. Moreover, the pore water on the bottom layer diffuses faster than on the top, because the top of the soil layer is poorly breathable silt, where the excess pore water dissipates more slowly. When $t=10^2$s, the maximum value of excess pore water pressure in the foundation is 149.4KPa, when $t=10^4$s, the maximum value of excess pore water pressure in the foundation is 127.3KPa, when $t=10^6$s, the maximum value of excess pore water pressure in the foundation is 49.2KPa. When $t=10^7$s, the excess pore water pressure tends to be 0KPa across the soil layer, and the pore water pressure in the foundation caused by the load is dissipated, which means that the drainage consolidation of soil is completed.

## 5. Discussion

### 5.1 Advantages of the Proposed Method

There are two advantages of the proposed deep learning approach.

(1) Compared with the traditional methods, the proposed method is computationally simple. In the proposed PINN method, the physical information such as partial differential equations, initial conditions, and boundary conditions of soil consolidation is defined without having actual data, and the data in the computational domain are randomly and automatically generated for model training.

(2) Compared with the traditional methods, the proposed method is computationally efficient. In the proposed PINN method, defined physical information is loosely coupled. For the prediction of excess pore water pressure in two-dimensional soil consolidation, it can be adjusted and highly adaptable to consolidation problems in different engineering environments without the need for re-modeling.

### 5.2 Shortcomings of the Proposed Method

The proposed method is proposed based on PINN. PINN has good performance in solving two-dimensional soil consolidation, but complex high-dimensional PDEs usually have no precise solutions，there is no completely accurate reference value to judge the training accuracy of the PINN training model. In addition, for complex engineering problems, stronger boundary conditions are needed to improve the fit to the engineering problems. Therefore, for some high-dimensional problems with weak boundary conditions, deep neural networks, as a general function approximator, can only obtain an approximate solution to the problem by minimizing the loss of the training model. How to further improve the computational accuracy is the current problem that PINN needs to be improved.

**5.3 Outlook and Future Work**

In future work, we will consider various constraints to improve the training accuracy of the neural network as much as possible and assess the applicability of this proposed deep learning approach using PINN to other engineering geology problems.

There are several strategies to improve the accuracy of neural network model training results. Adding more data is a good idea for ordinary models, but this idea does not apply to complex geological engineering problems with a lack of observation data. At present, the most effective method for PINN is algorithm optimization. It has been well understood that deep learning algorithms are driven by parameters that mainly affect the learning process results. In the future, we plan to combine several algorithms to achieve high-precision models, However, the choice of algorithm is difficult, this intuition comes from experience and practice, therefore, all relevant models should be applied and comparative performance checked.

The advantage of PINN in solving geotechnical problems is the use of prior knowledge and logic to discover the characteristics of the problem, but it lacks the ease of consistency with real data[32]. Further research is needed to achieve a perfect combination of physical knowledge and neural networks. Many improvements to the methods currently proposed are still possible, noteworthy, some theoretical problems remain unresolved. There is still a potential for development in optimizing training PINNs and expanding PINN to solve multiple equations.

We present a simple PINN problem for predicting the excess pore water pressure of two-dimensional consolidation, but this method can be extended to many large and complex multi-directional soil consolidation problems. The combination of deep learning models and physical laws is a new trend in the development of engineering geology, which is still in the initial stage of research and has not yet been widely applied to practical engineering. In the future, such as foundation deformation monitoring of large facilities, monitoring and early warning of landslides can be tried to be solved using the proposed deep learning method.

**6. Conclusion**

In this paper, we propose a deep learning method using PINN to predict the excess pore water pressure of two-dimensional soil consolidation. The essential idea behind the proposed method is to implement data-free model training with physics-informed constrained neural networks. In the proposed method, we present two simple examples of how to predict soil excess pore water pressure with different boundary conditions. In the proposed method, (1) a fully connected neural network is constructed, (2) the computational domain, partial differential equation (PDE), and constraints are defined to generate data for model training, and (3) the PDE of two-dimensional soil consolidation and the model of the neural network is connected to reduce the loss of the model. The effectiveness of the method is verified by comparing it with the numerical solution of the PDE for two-dimensional consolidation. The result indicates that: (1) the training result of the model is close to the numerical solution of two-dimensional consolidation, which verifies the validity of the method, (2) the more physical constraints there are, the closer the prediction is to the real value, (3) the excess pore water pressure of the soil can be predicted simply and efficiently, that provides great potential for real-time numerical prediction of a digital twin. In the future, the proposed deep learning approach can be used to investigate the large and complex multi-directional soil consolidation.

**Reference**

1	Terzaghi, K. *Erdbaumechanik auf bodenphysikalischer Grundlage*.  (Leipzig : Deuticke, 1925).
2	Biot, M. A. General Theory of Three‐Dimensional Consolidation. *Journal of Applied Physics* **12**, 155-164, doi:10.1063/1.1712886 (1941).
3	Schiffman, R. L. & Stein, J. R. One-Dimensional Consolidation of Layered Systems. *Journal of the Soil Mechanics and Foundations Division* **96**, 1499-1504, doi:10.1061/jsfeaq.0001453 (1970).
4	Wilson, N. E. & Elgohary, M. M. Consolidation of Soils Under Cyclic Loading. *Canadian Geotechnical Journal* **11**, 420-423, doi:10.1139/t74-042 (1974).
5	Zhu, G. & In, J. H. Y. J. G. Consolidation of double soil layers under depth-dependent ramp load. *Geotechnique* **49**, págs. 415-421 (1999).
6	Indraratna, B., Rujikiatkamjorn, C. & Sathananthan, I. Radial consolidation of clay using compressibility indices and varying horizontal permeability. *Canadian Geotechnical Journal* **42**, 1330-1341, doi:10.1139/t05-052 (2005).
7	Lo, W.-C., Sposito, G., Lee, J.-W. & Chu, H. One-dimensional consolidation in unsaturated soils under cyclic loading. *Adv Water Resour* **91**, 122-137, doi:10.1016/j.advwatres.2016.03.001 (2016).
8	Pham, B. T. *et al.* A novel artificial intelligence approach based on Multi-layer Perceptron Neural Network and Biogeography-based Optimization for predicting coefficient of consolidation of soil. *Catena* **173**, 302-311, doi:10.1016/j.catena.2018.10.004 (2019).
9	Yarushina, V. M. & Podladchikov, Y. Y. (De)compaction of porous viscoelastoplastic media: Model formulation. *Journal of Geophysical Research: Solid Earth* **120**, 4146-4170, doi:10.1002/2014jb011258 (2015).
10	Ho, L., Fatahi, B. & Khabbaz, H. Analytical solution for one-dimensional consolidation of unsaturated soils using eigenfunction expansion method. *International Journal for Numerical and Analytical Methods in Geomechanics* **38**, 1058-1077, doi:10.1002/nag.2248 (2014).
11	Cavalcanti, M. C. & Telles, J. C. F. Biot's consolidation theory—application of BEM with time independent fundamental solutions for poro-elastic saturated media. *Engineering Analysis with Boundary Elements* **27**, 145-157, doi:10.1016/s0955-7997(02)00092-9 (2003).
12	Vilà, J., González, C. & Llorca, J. A level set approach for the analysis of flow and compaction during resin infusion in composite materials. *Composites Part A: Applied Science and Manufacturing* **67**, 299-307, doi:10.1016/j.compositesa.2014.09.002 (2014).

13    Wang, L., Sun, D. a., Li, L., Li, P. & Xu, Y. Semi-analytical solutions to one-dimensional consolidation for unsaturated soils with symmetric semi-permeable drainage boundary. *Computers and Geotechnics* **89**, 71-80, doi:10.1016/j.compgeo.2017.04.005 (2017).

14    Lu, L., Meng, X., Mao, Z. & Karniadakis, G. E. DeepXDE: A Deep Learning Library for Solving Differential Equations. *Siam Rev* **63**, 208-228, doi:10.1137/19m1274067 (2021).

15    Raissi, M., Perdikaris, P. & Karniadakis, G. E. Physics-informed neural networks: A deep learning framework for solving forward and inverse problems involving nonlinear partial differential equations. *Journal of Computational Physics* **378**, 686-707, doi:10.1016/j.jcp.2018.10.045 (2019).

16    Blechschmidt, J. & Ernst, O. G. Three ways to solve partial differential equations with neural networks — A review. *GAMM-Mitteilungen* **44**, e202100006, doi:10.1002/gamm.202100006 (2021).

17    Owhadi, H. Bayesian Numerical Homogenization. *Multiscale Modeling & Simulation* **13**, 812-828, doi:10.1137/140974596 (2015).

18    Raissi, M., Perdikaris, P. & Karniadakis, G. E. Inferring solutions of differential equations using noisy multi-fidelity data. *Journal of Computational Physics* **335**, 736-746, doi:10.1016/j.jcp.2017.01.060 (2017).

19    Raissi, M., Perdikaris, P. & Karniadakis, G. E. Machine learning of linear differential equations using Gaussian processes. *Journal of Computational Physics* **348**, 683-693, doi:10.1016/j.jcp.2017.07.050 (2017).

20    Karniadakis, G. E. *et al.* Physics-informed machine learning. *Nature Reviews Physics* **3**, 422-440, doi:10.1038/s42254-021-00314-5 (2021).

21    Raissi, M., Yazdani, A. & Karniadakis, G. E. Hidden fluid mechanics: Learning velocity and pressure fields from flow visualizations. *Science* **367**, 1026-1030, doi:10.1126/science.aaw4741 (2020).

22    Tartakovsky, A. M., Marrero, C. O., Perdikaris, P., Tartakovsky, G. D. & Barajas‐Solano, D. Physics‐Informed Deep Neural Networks for Learning Parameters and Constitutive Relationships in Subsurface Flow Problems. *Water Resources Research* **56**, doi:10.1029/2019wr026731 (2020).

23    Almajid, M. M. & Abu-Al-Saud, M. O. Prediction of porous media fluid flow using physics informed neural networks. *Journal of Petroleum Science and Engineering* **208**, doi:10.1016/j.petrol.2021.109205 (2022).

24    Depina, I., Jain, S., Mar Valsson, S. & Gotovac, H. Application of physics-informed neural networks to inverse problems in unsaturated groundwater flow. *Georisk: Assessment and Management of Risk for Engineered Systems and Geohazards* **16**, 21-36, doi:10.1080/17499518.2021.1971251 (2021).

25    Bandai, T. & Ghezzehei, T. A. Physics‐Informed Neural Networks With Monotonicity Constraints for Richardson‐Richards Equation: Estimation of Constitutive Relationships and Soil Water Flux Density From Volumetric Water

Content Measurements. *Water Resources Research* **57**, e2020WR027642, doi:10.1029/2020wr027642 (2021).

26  Zhang, Z. A physics-informed deep convolutional neural network for simulating and predicting transient Darcy flows in heterogeneous reservoirs without labeled data. *Journal of Petroleum Science and Engineering* **211**, doi:10.1016/j.petrol.2022.110179 (2022).

27  Bekele, Y. W. Physics-informed deep learning for one-dimensional consolidation. *Journal of Rock Mechanics and Geotechnical Engineering* **13**, 420-430, doi:10.1016/j.jrmge.2020.09.005 (2021).

28  Waheed, U. b., Haghighat, E., Alkhalifah, T., Song, C. & Hao, Q. PINNeik: Eikonal solution using physics-informed neural networks. *Computers & Geosciences* **155**, 104833, doi:10.1016/j.cageo.2021.104833 (2021).

29  Cuomo, S. *et al.* Scientific Machine Learning through Physics-Informed Neural Networks: Where we are and What's next. arXiv:2201.05624 (2022). <https://ui.adsabs.harvard.edu/abs/2022arXiv220105624C>.

30  L.Rendulic. Porenziffer und Porenwasserdruck in Tonen. *Der Bauingenieur* **17**, 559-564 (1936).

31  Yan, S., W., Chu & Improvement, J. J. G. Soil improvement for a road using the vacuum preloading method. (2003).

32  Garnelo, M. & Shanahan, M. Reconciling deep learning with symbolic artificial intelligence: representing objects and relations. *Current Opinion in Behavioral Sciences* **29**, 17-23, doi:10.1016/j.cobeha.2018.12.010 (2019).